\documentclass[twocolumn,showpacs,pra]{revtex4}

\usepackage{amsmath,subfigure,psfrag,color}
\def\U#1{{\rm #1}} 
\def\u#1{_{\rm #1}}
\newcommand{\od}[2]{\frac{\mathrm{d} #1}{\mathrm{d} #2}}

\newcommand{\del}{\partial}

\newcommand{\vect}[1]{\boldsymbol{#1}}
\newcommand{\bra}[1]{\langle #1 |}
\newcommand{\ket}[1]{| #1 \rangle}

\newcommand{\dagg}[1]{#1 ^\dagger}
\newcommand{\inner}[2]{\langle #1 , #2 \rangle}
\def\W{{\rm W}}

\def\H{{\rm H}}
\def\V{{\rm V}}

\def\d{\U{d}}

\begin{document}
\title{
Optimal local expansion of W states using linear optics and Fock states
}
\author{Rikizo Ikuta}
\affiliation{Graduate School of Engineering Science, Osaka University, Toyonaka, Osaka 560-8531, Japan}
\author{Toshiyuki Tashima}
\affiliation{Graduate School of Engineering Science, Osaka University, Toyonaka, Osaka 560-8531, Japan}
\author{Takashi Yamamoto}
\affiliation{Graduate School of Engineering Science, Osaka University, Toyonaka, Osaka 560-8531, Japan}
\author{Masato Koashi}
\affiliation{Graduate School of Engineering Science, Osaka University, Toyonaka, Osaka 560-8531, Japan}
\author{Nobuyuki Imoto}
\affiliation{Graduate School of Engineering Science, Osaka University, Toyonaka, Osaka 560-8531, Japan}

\begin{abstract}
We derive 
the maximum success probability of circuits 
with passive linear optics 
for post-selectively 
expanding an $N$-photon W state 
to an $(N+n)$-photon W state, 
by accessing only one photon of the initial W state 
and adding $n$ photons in a Fock state. 
We show that the maximum success probability is achieved 
by a polarization-dependent beamsplitter 
and $n-1$ polarization-independent beamsplitters.
\end{abstract}
\maketitle
\section{Introduction}
\label{sec1}

Multipartite entanglement has rich structure and diverse properties, 
which can be exploited for many kinds of applications. 
With an increasing number of parties sharing entanglement, 
the fact that the structure becomes more complex also 
means that preparation of such an entanglement may 
become more difficult, 
especially if one tries to generate the whole state 
in a single step. 
Instead, 
one may start with an initial entangled state 
among a small number of subsystems, 
and then expand it by adding ancillary subsystems. 
In fact, 
many interesting classes of multipartite entangled states, 
such as 
Greenberger-Horne-Zeilinger (GHZ) states \cite{GHZ}, 
cluster states \cite{cluster} 
and W states \cite{durW}, 
are defined for an arbitrary number of qubits, 
which makes the expansion strategy look attractive. 
There are several proposals of state expansion 
for multipartite entangled states 
by local manipulation on a single site without 
accessing other qubits 
\cite{tashima1, gong, tashima2}. 
For example, in the case of GHZ states, 
a deterministic local expansion 
of an $N$-qubit state to an $(N+n)$-qubit state 
is possible in principle. 
In optical systems, several experimental demonstrations 
have been performed 
\cite{GHZ1, GHZ2, GHZ3, GHZ4} 
by using quantum parity checking gates~\cite{pittman}. 

W states are an interesting class of 
multipartite entangled states 
in that 
they have a web-like entanglement structure. 
An $N$-qubit W state is represented by 
$\ket{{\rm W}_N}=(\ket{10\cdots 0}
+\ket{010\cdots 0}+\cdots +\ket{0\cdots 01})/\sqrt{N}$. 
Every qubit in $\ket{\W_N}$ has bonds with every other qubit, 
and the pairwise entanglement survives 
even if all the other ($N-2$)-qubits are discarded 
\cite{durW, koashiW, liuW}. 
In recent years, 
there have been a number of theoretical proposals and 
experimental demonstrations using W states 
in multiparty protocols 
such as quantum key distribution \cite{jooQKDW}, leader election \cite{hondtLE} 
as well as preparation of W states in optical systems 
\cite{yamamotoW, zouW, kieselW, liW, eiblW, limW, mikamiW, nhaW,tomitaW,kimbleW,tashima3}. 
In the case of W states, 
a deterministic local expansion is 
impossible even in principle 
because the marginal state of the 
remaining untouched $N-1$ qubits is different 
for $\ket{{\rm W}_N}$ and $\ket{{\rm W}_{N+n}}$, 
so it is worth seeking nontrivial ways 
of expanding W states probabilistically. 
Since the qualitative difference from GHZ states 
in the expandability may arise from 
the difference in the nature of multipartite entanglement, 
study of efficient local expansions of W states 
is interesting theoretically as well as practically. 

For photonic polarization-based qubits, 
recent proposals and a demonstration include 
the expansion of $\ket{{\rm W}_N}$ 
to $\ket{{\rm W}_{N+1}}$ \cite{tashima1, gong} 
and to $\ket{{\rm W}_{N+2}}$ \cite{tashima2, tashima4}. 
These expansion schemes are composed of passive linear optics and 
one or two photons in a Fock state. 
In this paper, 
we address the question of what is the best way of 
expanding $\ket{\W_N}$ 
to $\ket{\W_{N+n}}$ for photonic polarization-based qubits. 
We discuss the maximum success probability of local expansion methods 
composed of passive linear optics and an ancilla mode in an $n$-photon Fock state. 
We derive the maximum success probability, 
and also show that it is achieved by a 
polarization-dependent beamsplitter (PDBS) 
and $n-1$ polarization-independent beamsplitters (BSs).
In the case of $n=2$, 
the optimal success probability is higher 
than that of the expanding gate proposed so far \cite{tashima2}. 

This paper is organized as follows: 
In Sec.~\ref{sec2}, we describe 
the expansion schemes considered in this paper. 
In Sec.~\ref{sec3}, we derive an upper bound on 
the success probability for expanding W states. 
In Sec.~\ref{sec4}, we explicitly construct 
a linear optical circuit that achieves 
the derived upper bound on the success probability. 
Finally, in Sec.~\ref{sec5}, we give a brief summary and conclusions.

\section{Expansion methods of W states}
\label{sec2}

\begin{figure}[t]
 \begin{center}
 \scalebox{1.0}{\includegraphics{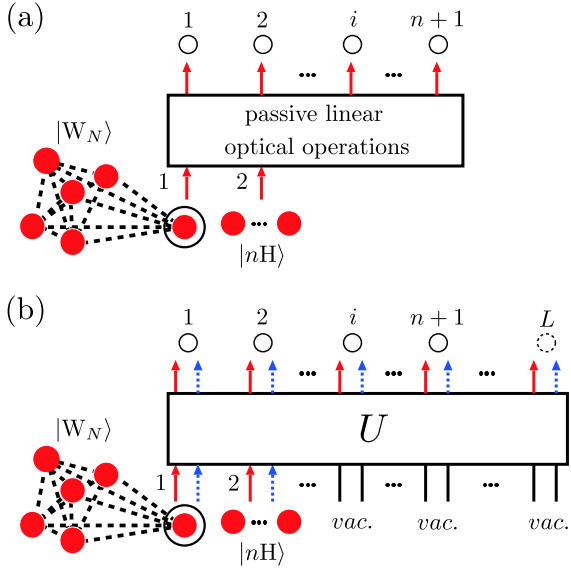}}
  \caption{(Color online) Local expansion circuits of $\ket{\W_N}$ 
  to $\ket{\W_{N+n}}$. 
  (a) The expansion circuit considered in this paper. 
  This circuit is composed of only passive linear optics. 
  (b) The expansion circuit in Fig. \ref{fig:ExGate}a 
  is equivalent to a lossless linear optical circuit like this figure. 
  Solid(dashed) arrows mean H(V)-polarized modes. 
  \label{fig:ExGate}}
 \end{center}
\end{figure}
The optical circuit considered in this paper 
for expanding W states 
is composed of 
only passive linear optics, that is, of 
PDBSs, BSs, phase shifters, wave plates, 
and arbitrary linear losses. 
As shown in Fig.~\ref{fig:ExGate}~(a), 
the circuit has two input spacial modes 1 and 2. 
One photon from an $N$-photon polarization-entangled W state, 
which is represented by 
$\ket{\W_N}=
(\ket{\V\H\cdots \H}+\ket{\H\V\H\cdots \H}+\cdots 
+\ket{\H\cdots \H\V})/\sqrt{N}
$ 
where $\ket{\H(\V)}$ is the state of an 
horizontally (H-) and vertically (V-) 
polarized photon, 
is fed to mode 1, 
and an H-polarized $n$-photon Fock state 
(we denote it by $\ket{n\H}$) 
is fed to mode 2. 
We define the successful operation of 
the circuit to be the events 
where the photons come out 
from the output spacial modes $1,\ldots, n+1$, one by one, 
namely, exactly one photon from every output mode. 
We require that the output state is exactly 
the $(N+n)$-photon W state for the success events. 

As is well known, linear optical losses can be equivalently 
described by lossless optical circuits with auxiliary spacial modes. 
This is because a linear loss with transmission $T$ is 
equivalently realized by using a BS with transmission $T$. 
Since a BS is a lossless component with two input modes 
and two output modes, we can simulate losses with lossless 
components by introducing auxiliary input modes initially 
in the vacuum and the same number of auxiliary output modes 
from which the `lost' photons escape. 
The circuit in Fig.~\ref{fig:ExGate}~(a) is thus 
described equivalently by a lossless circuit with 
$L(\geq n+1)$ input and output spacial modes. 
The family of schemes considered in this paper is defined as follows. 
The expansion circuit has $2L$ input and output modes 
composed of $L$ spatial modes 
with each having two polarizing 
(H- and V-polarized) modes. 
One photon from an $N$-photon polarization-entangled 
W state enters into input spacial mode $1$ and 
an H-polarized $n$-photon Fock state 
enters into input spacial mode $2$. 
All of the other input spacial modes receive 
vacuum states. 
Arbitrary lossless linear optical operations are 
applied to the $2L$ input modes. 
We show this expansion circuit in Fig.~\ref{fig:ExGate}~(b).

Let $\ket{vac}\u{in(out)}$ be the state of 
all the $2L$ input(output) modes in the vacuum. 
The unitary operation $U$ for 
any passive linear optical operations satisfies 
$U\ket{vac}\u{in}=\ket{vac}\u{out}$, and 
relevant actions of $U$ are represented by 
\begin{eqnarray}
U\dagg{a}_{1\H}\dagg{U}&=&
\sum^{L}_{j=1}
(\beta_{j\H}\dagg{b}_{j\H}
+\beta_{j\V}\dagg{b}_{j\V})\ ,
\label{a1}\\
U\dagg{a}_{1\V}\dagg{U}&=&
\sum^{L}_{j=1}
(\gamma_{j\H}\dagg{b}_{j\H}
+\gamma_{j\V}\dagg{b}_{j\V})\ , 
\label{a2}
\end{eqnarray}
and
\begin{eqnarray}
U \dagg{a}_{2\H}\dagg{U}&=&
\sum^{L}_{j=1}
(\alpha_{j\H}\dagg{b}_{j\H}
+\alpha_{j\V}\dagg{b}_{j\V})\ , 
\label{a3}
\end{eqnarray}
where 
$\dagg{a}_{1\H(\V)}$ and $\dagg{a}_{2\H(\V)}$ 
are the H(V)-polarized photon creation operators 
for input spacial mode 1 and 2, 
$\dagg{b}_{j\H(\V)}$ is the H(V)-polarized photon creation operator 
for output spacial mode $j$, and 
$\alpha_{j\H(\V)}$, $\beta_{j\H(\V)}$ and $\gamma_{j\H(\V)}$ are 
complex numbers. 
Using the commutation relation 
$[a_{js},\dagg{a_{j's'}}]=
[b_{js},\dagg{b_{j's'}}]=\delta_{jj'}\delta_{ss'}(s,s'=\H,\V)$ 
and Eqs.~(\ref{a1}) -- (\ref{a3}), 
we obtain 
\begin{eqnarray}
\sum_{j=1}^{L}
(\Omega_{j\H}\Omega'^{*}_{j\H}+\Omega_{j\V}\Omega'^{*}_{j\V})
=\delta_{\Omega\Omega'}, 
\label{commutation}
\end{eqnarray}
for $\Omega, \Omega' =\alpha,\beta,\gamma$. 
By denoting $\ket{\H_{N}}\equiv \ket{\H\cdots \H}$, 
the input state to the expansion circuit 
in Fig.~\ref{fig:ExGate} can be written as 
\begin{eqnarray}
\ket{\W_{N}} \ket{n\H}
&\hspace{-0.2cm}
=
\hspace{-0.2cm}&
\frac{1}{\sqrt{N}}
\left(
\sqrt{N-1}\ket{\W_{N-1}}\otimes 
\frac{\dagg{a\u{1H}}(\dagg{a\u{2H}})^n}{\sqrt{n!}}
\ket{vac}\u{in}\right.\nonumber \\
&\hspace{-0.2cm}
+
\hspace{-0.2cm}&
\left. 
\ket{\H_{N-1}}\otimes 
\frac{\dagg{a\u{1V}}(\dagg{a\u{2H}})^n}{\sqrt{n!}}
\ket{vac}\u{in}
\right), 
\label{initial}
\end{eqnarray}
while state $\ket{\W_{N+n}}$ which we desire as an output 
is 
\begin{eqnarray}
\ket{\W_{N+n}}
&\hspace{-0.2cm}=
\hspace{-0.2cm}&
\frac{1}{\sqrt{N+n}}
\left(
\sqrt{N-1}\ket{\W_{N-1}}\otimes 
\prod_{j=1}^{n+1}\dagg{b_{j\H}}\ket{vac}\u{out}
\right.\nonumber\\
&\hspace{-0.2cm}&
\hspace{-0.5cm}
+
\left. 
\ket{\H_{N-1}}\otimes 
\sum_{i=1}^{n+1}\dagg{b_{i\V}}\prod_{j\neq i}^{n+1}\dagg{b_{j\H}}\ket{vac}\u{out}
\right). 
\label{final}
\end{eqnarray}
The post-selected events where the photons come out 
from output spacial mode $1,\ldots, n+1$ one by one are 
described by a projector written as 
\begin{eqnarray}
\Pi\u{post}&\equiv &
\sum_{s_1=\H,\V}\cdots \sum_{s_{n+1}= \H,\V}
\nonumber\\
&&
\hspace{0.5cm}
\left(
\prod_{l=1}^{n+1}
\dagg{b}_{ls_l}
\right)
\ket{vac}\bra{vac}\u{out}
\left(
\prod_{l'=1}^{n+1}
b_{l's_{l'}}
\right).
\label{projection}
\end{eqnarray}
By applying $U$ and $\Pi\u{post}$ 
to Eq.~(\ref{initial}), 
we should obtain Eq.~(\ref{final}). Hence 
\begin{eqnarray}
\hspace{-0.4cm}
\Pi\u{post} U \ket{\W_{N}} \ket{n\H}
=\sqrt{P\u{suc}}\ket{\W_{N+n}}, 
\label{postselect}
\end{eqnarray}
where $P\u{suc}$ is the probability of success. 
By operating 
$\bra{\W_{N-1}}\otimes \bra{vac}\u{out}(\prod_{l=1}^{n+1}b_{l\H})$ 
on Eq.~(\ref{postselect}) from the left, 
we have 
\begin{eqnarray}
\sqrt{P\u{suc}} = \sqrt{\frac{n!(N+n)}{N}}\eta_0\ , 
\label{successroot0}
\end{eqnarray}
where 
\begin{eqnarray}
\eta_0 &\equiv&
\bra{vac}\u{out}
\left(
\prod_{l=1}^{n+1}b_{l\H}
\right)
U
\frac{\dagg{a\u{1\H}}(\dagg{a\u{2H}})^n}{n!}
\ket{vac}\u{in}\\
&=&
\sum_{i=1}^{n+1}\beta_{i\H}
\prod_{j\neq i}^{n+1}\alpha_{j\H}\ . 
\label{eta0}
\end{eqnarray}
Here we have used 
Eqs.~(\ref{a1})--(\ref{a3}), 
(\ref{initial}) and (\ref{final}). 
Similarly, by operating 
$\bra{\H_{N-1}}\otimes \bra{vac}\u{out}(b_{i\V}\prod_{l\neq i}^{n+1}b_{l\H})$ 
on Eq.~(\ref{postselect}) from the left, 
we have 
\begin{eqnarray}
\sqrt{P\u{suc}} = \sqrt{\frac{n!(N+n)}{N}}\eta_i\ , 
\label{successrooti}
\end{eqnarray}
where 
\begin{eqnarray}
\eta_i&\equiv&
\bra{vac}\u{out}
\left(
b_{i\V}
\prod_{l\neq i}^{n+1}b_{l\H}
\right)
U
\frac{\dagg{a\u{1V}}(\dagg{a\u{2H}})^n}{n!}
\ket{vac}\u{in}
\label{etai0}\\ 
&=&
\gamma_{i\V}\prod_{j\neq i}^{n+1}
\alpha_{j\H}
+
\alpha_{i\V} 
\sum_{j\neq i}^{n+1}
\gamma_{j\H}
\prod_{k\neq i,j}^{n+1}
\alpha_{k\H}\ , 
\label{etai}
\end{eqnarray}
for $1\leq i\leq n+1$. 
Here we have used 
Eqs.~(\ref{a1})--(\ref{a3}), 
(\ref{initial}) and (\ref{final}). 
From Eqs.~(\ref{successroot0}) and (\ref{successrooti}), 
we obtain 
\begin{eqnarray}
P\u{suc} = \frac{n!(N+n)}{N }|\eta_0|^2\ ,
\label{success}
\end{eqnarray}
and 
\begin{eqnarray}
\eta_0 = \cdots =\eta_{n+1}\ . 
\label{eta}
\end{eqnarray}

\section{Bound on the success probability}
\label{sec3}

In this section, 
we derive an upper bound on success probability $P\u{suc}$. 
Since the cases with $P\u{suc}=0$ are irrelevant 
for the upper bound, 
we focus on the cases with $P\u{suc}>0$ here. 
The requirement of producing the exact W state, Eq.~(\ref{postselect}), 
narrows down the choices of the unitary operator $U$ significantly. 
As shown in Appendix \ref{appA}, 
in order to satisfy 
$P\u{suc}> 0$, 
it is necessary that 
$\alpha_{i\V}=\beta_{i\V}=0$ for $1\leq i\leq n+1$ 
and $\prod_{j=1}^{n+1}\alpha_{j\H}\neq 0$. 
Thus, from Eqs.~
(\ref{eta0}) and (\ref{etai}), 
we are allowed to express $\eta_0$ and $\eta_i$, for later use, 
as 
\begin{eqnarray}
\eta_0&=&
\left( 
\prod_{i=1}^{n+1}
\alpha_{i\H}
\right)
\sum_{j=1}^{n+1}
\frac{\beta_{j\H}}{\alpha_{j\H}},
\label{eta02}\\
\eta_i&=& \gamma_{i\V}
\prod_{j\neq i}^{n+1}\alpha_{j\H}, 
\label{etai2}
\end{eqnarray}
for $1\leq i\leq n+1$. 
Define $P_i\equiv |\alpha_{i\H}|^2$ ($1\leq i\leq n+1$), 
$\vect{P}\equiv (P_1,\ldots ,P_{n+1})$, 
\begin{eqnarray}
S(\vect{P})\equiv \sum_{j=1}^{n+1}P_j\ , 
\label{SP}
\end{eqnarray}
and 
\begin{eqnarray}
\Pi(\vect{P})\equiv \prod_{j=1}^{n+1}P_j\ .
\label{PI}
\end{eqnarray}
Because $\prod_{j=1}^{n+1}\alpha_{j\H}\neq 0$ 
and 
$S(\vect{P})\leq 1$ from Eq.~(\ref{commutation}), 
we obtain 
\begin{eqnarray}
\hspace{-0.4cm}
\vect{P}\in \mathcal{R}\equiv 
\{ \vect{P}| P_i > 0 (1\leq i\leq n+1),0 < S(\vect{P})\leq 1\}. 
\label{PinR}
\end{eqnarray}
Combining $\alpha_{i\V}=\beta_{i\V}=0$ 
for $1\leq i\leq n+1$ 
with 
$\sum_{j=1}^{L}
(\alpha_{j\H}^*\beta_{j\H}
+\alpha_{j\V}^*\beta_{j\V})=0$ 
and 
$\sum_{j=1}^{L}(|\alpha_{j\H}|^2+|\alpha_{j\V}|^2)=
\sum_{j=1}^{L}(|\beta_{j\H}|^2+|\beta_{j\V}|^2)=1$ 
from Eq.~(\ref{commutation}), 
we have 
\begin{eqnarray}
\left| \sum_{j=1}^{n+1}\frac{\beta_{j\H}}{\alpha_{j\H}}\right|^2
&=&\left| \sum_{j=1}^{L}
\sum_{s=\H,\V}
(
\zeta_{js}-\alpha_{js}^*(n+1)
)\beta_{js}\right|^2\nonumber\\
&\leq &
\sum_{j=1}^{L}
\sum_{s=\H,\V}
\left| \zeta_{js}-\alpha_{js}^*(n+1)\right|^2\nonumber\\
&=&\sum_{j=1}^{n+1}\frac{1}{P_j}-(n+1)^2, 
\label{Cauchy}
\end{eqnarray}
where 
$\zeta_{i\H}\equiv \alpha_{i\H}^{-1}$ 
and $\zeta_{i\V}\equiv 0$ 
for $1\leq i\leq n+1$, and 
$\zeta_{i\H} = \zeta_{i\V}\equiv 0$ 
for $n+2\leq i\leq L$. 
Here we have used 
the Cauchy-Schwarz inequality. 
From Eqs.~(\ref{eta02}), (\ref{PI}) and (\ref{Cauchy}), 
we obtain a bound on $|\eta_0|^2$ written as 
\begin{eqnarray}
|\eta_0|^2\leq 
F(\vect{P})\equiv 
\Pi(\vect{P})
\left(
\sum_{k=1}^{n+1}\frac{1}{{P_k}} - (n+1)^2
\right). 
\label{boundF}
\end{eqnarray}
Note that $F(\vect{P}) > 0$ 
since we are focusing on the cases with $P\u{suc} > 0$. 
Using Eqs.~(\ref{eta}), (\ref{etai2}), (\ref{SP}) and 
$\sum_{j=1}^{n+1}|\gamma_{j\V}|^2\leq 1$ 
from Eq.~(\ref{commutation}), 
we have 
$|\eta_0|^2S(\vect{P})
=\sum_{j=1}^{n+1}P_j |\eta_j|^2
=\sum_{j=1}^{n+1}P_j|\gamma_{j\V}|^2\prod_{k\neq j}^{n+1}P_k
=\sum_{j=1}^{n+1}|\gamma_{j\V}|^2
\Pi(\vect{P})
\leq \Pi(\vect{P})$. 
Then we obtain another bound on $|\eta_0|^2$ written as 
\begin{eqnarray}
|\eta_0|^2\leq 
G(\vect{P})\equiv \frac{\Pi(\vect{P})}{S(\vect{P})}. 
\label{boundG}
\end{eqnarray}
From Eqs.~(\ref{success}), (\ref{PinR}), (\ref{boundF}) and (\ref{boundG}), 
we obtain 
\begin{eqnarray}
P\u{suc} \leq \frac{n!(N+n)}{N} 
\max_{\vect{P}\in \mathcal{R}} H(\vect{P})\ , 
\label{success2}
\end{eqnarray}
where $H(\vect{P})\equiv \min\{ F(\vect{P} ),G(\vect{P}) \}$. 

Before conducting the optimization over $\vect{P}$ 
in Eq.~(\ref{success2}), 
let us discuss physical intuition behind the bound $H(\vect{P})$. 
Recall that $P_i=|\alpha_{i\H}|^2$ is the probability 
of a photon in the input mode $2$ in Fig.~\ref{fig:ExGate}~(a) 
to appear at the output mode $i$. 
One of the bounds on $|\eta_0|^2$, $G(\vect{P})$, 
was derived through constraints on 
$|\eta_i|^2$, which is proportional to the probability of having a
V-polarized photon at the output mode $i$ 
and $n$ H-polarized photons in the other $n$ output modes, 
one in each~[See Eq.~(\ref{etai0})]. 
From the definition of $G(\vect{P})$ in Eq.~(\ref{boundG}), 
we see that
\begin{eqnarray}
S(\vect{P})=P_1+ \cdots +P_{n+1}=1
\label{lossless}
\end{eqnarray}
and
\begin{eqnarray}
P_1=\cdots =P_{n+1}
\label{equal}
\end{eqnarray}
give the maximum of $G(\vect{P})$. 
This means lossless and equal distribution of 
the $n$ H-polarized photons incident on the input
mode $2$ is the best for maximizing the amplitude of the terms
including one V-polarized photon. 
This result does not change 
even if we regard photons as classical distinguishable particles, 
since the origin of the V photon (input mode $1$) 
and that of an H photon (input mode $2$) are uniquely determined 
and no interference occurs. 

On the other hand, the bound $F(\vect{P})$ stems directly 
from a constraint on $|\eta_0|^2$, 
which is proportional to the probability 
of $(n+1)$ H-polarized photons to appear at the $(n+1)$ output modes, 
one by one. 
In this case, there are $(n+1)$ indistinguishable paths, 
depending on which of the $(n+1)$ output photons 
is traced back to the input photon in mode $1$. 
As a result, the total amplitude $\eta_0$ 
is given by the sum over $(n+1)$ terms as in Eq.~(\ref{eta0}), 
which can be rewritten as 
\begin{eqnarray}
\eta_0 = \left( \prod_{j=1}^{n+1} \alpha_{j\H}\right) 
\sum_{i=1}^{n+1} \frac{1}{P_i} \alpha_{i\H}^* \beta_{i\H}\ . 
\label{eta0re}
\end{eqnarray}
Let us see the interference among these terms at the choice of 
$\vect{P}$ satisfying Eqs.~(\ref{lossless}) and (\ref{equal}). 
The parameters $\beta_{i\H} (i=1, \ldots, L)$ describe 
how the input photon in mode $1$ is distributed. 
They must satisfy the unitarity condition of Eq.~(\ref{commutation}), 
which gives, under the lossless condition of Eq.~(\ref{lossless}), 
a constraint 
\begin{eqnarray}
\sum_{i=1}^{n+1} \alpha_{i\H}^* \beta_{i\H}=0\ .
\label{alphabeta}
\end{eqnarray}
Together with Eq.~(\ref{equal}), 
we see that the total amplitude $\eta_0$ 
always vanishes regardless of the choice of the parameters 
$\beta_{i\H} (i=1,\ldots ,n+1)$. 
{This is the reason why we have 
$F(\vect{P})\rightarrow 0$ for 
$\vect{P}\rightarrow ((n+1)^{-1},\ldots ,(n+1)^{-1})$. 
Incidentally, 
the case with $n=1$ is equivalent to the well-known two-photon 
interference effect at the symmetric beamsplitter \cite{mandel}, 
in which the two photons never leave separated. 
The above result with general $n$ can thus be regarded 
as an extension of the $(1,1)$-photon case to the $(1,n)$-photon case. 
What is interesting here is that 
the symmetry is required only for the $n$-photon input, 
and not for the one-photon input, 
to achieve the complete destructive interference. 

In order to obtain a nonzero value $H(\vect{P})>0$, 
one must go away from the point $P_1=\cdots =P_{n+1}=(n+1)^{-1}$ 
by dropping either the lossless condition Eq.~(\ref{lossless}) 
and/or the symmetry condition Eq.~(\ref{equal}). 
If one breaks the symmetry, 
the orthogonality condition Eq.~(\ref{alphabeta}) 
no longer implies $\eta_0=0$ in Eq.~(\ref{eta0re}). 
On the other hand, introduction of loss relaxes 
the orthogonality condition Eq.~(\ref{alphabeta}) itself, 
since it is equivalent to introducing auxiliary modes 
in Fig.~\ref{fig:ExGate}~(b), namely, $L>n+1$. 
The condition Eq.~(\ref{alphabeta}) then changes to 
\begin{eqnarray}
\sum_{i=1}^{n+1} \alpha_{i\H}^* \beta_{i\H}= 
-\sum_{i=n+2}^{L} \alpha_{i\H}^* \beta_{i\H}
-\sum_{i=n+2}^{L} \alpha_{i\V}^* \beta_{i\V}\ , 
\end{eqnarray}
which allows more freedom in the choice of parameters $\beta_{i\H}
(i=1,\ldots, n+1)$.

In the following, 
we optimize over $\vect{P}$ in Eq.~(\ref{success2}) by 
deriving necessary conditions for $\vect{P}$ 
to achieve the maximum of $H(\vect{P})$. 
We assume $P_1\leq \cdots \leq P_{n+1}$ 
without loss of generality, and 
we consider the two cases, 
$S(\vect{P}) < 1$ and 
$S(\vect{P}) = 1$, separately. 

In the case of $S(\vect{P}) < 1$, 
for $\vect{P}$ to be a local maximum, 
there exists $\epsilon >0$ such that 
$H(\vect{P}+\Delta \vect{P}_0) \leq 
H(\vect{P})$ 
for any 
$\Delta \vect{P}_0\equiv \alpha \vect{u}_0+\beta \vect{v}_0$ 
with $\alpha^2+\beta^2 < \epsilon$, 
where 
$\vect{u}_0 \equiv (1,0,\ldots, 0)$ and 
$\vect{v}_0 \equiv (0,\ldots, 0,1)$. 
As shown in Appendix \ref{appB}, 
this leads to 
\begin{eqnarray}
\inner{\vect{u}_0}{{\boldsymbol \nabla} F}
\inner{\vect{v}_0}{{\boldsymbol \nabla} G}
=
\inner{\vect{v}_0}{{\boldsymbol \nabla} F}\inner{\vect{u}_0}{{\boldsymbol \nabla} G}\ , 
\label{u0v0}
\end{eqnarray}
where $\inner{\vect{X}}{\vect{Y}}$ 
means the inner product between $\vect{X}$ and $\vect{Y}$. 
From 
$
\nabla_i F(\vect{P})=\del{F(\vect{P})}/\del{P_i}
=P_i^{-1}(F(\vect{P})-\Pi(\vect{P})P_i^{-1}), 
\del{G(\vect{P})}/\del{P_i}=G(\vect{P})(P_i^{-1}-S(\vect{P})^{-1})
$
and Eq.~(\ref{u0v0}), 
we obtain 
\begin{eqnarray}
\left(
\frac{1}{P_{n+1}}-\frac{1}{P_1}
\right)
\left( 
\frac{F(\vect{P})}{\Pi(\vect{P})}
 +\frac{1}{P_1P_{n+1}}
\sum_{j=2}^{n}P_j
\right)=0\ . 
\label{u0v02}
\end{eqnarray}
Since 
$P_1\leq \cdots \leq P_{n+1}$, 
Eq.~(\ref{u0v02}) means 
$P_1=\cdots =P_{n+1}=S(\vect{P}) (n+1)^{-1}$. 
At this point, 
$F(\vect{P})$ and $G(\vect{P})$ are regarded as 
functions of single parameter $S(\vect{P})$. 
We show in Appendix \ref{appC} 
that 
$S(\vect{P})=1-(n+1)^{-2}$ gives 
the local maximum of $H(\vect{P})$, 
whose value is 
\begin{eqnarray}
H\u{lossy}\equiv 
\frac{n^n(n+2)^n}{(n+1)^{3n+1}}\ . 
\label{Hlossy}
\end{eqnarray}

In the case of $S(\vect{P})=1$, 
since $P_i>0$ for all $i$, 
for $\vect{P}$ to be a local maximum 
under the constraint 
$S(\vect{P})=1$, 
there exists $\epsilon >0$ such that 
$H(\vect{P}+\Delta \vect{P}_1) \leq 
H(\vect{P})$ for any 
$\Delta \vect{P}_1\equiv \alpha \vect{u}_1+\beta \vect{v}_1$ 
with $\alpha^2+\beta^2 < \epsilon$, 
where 
$\vect{u}_1\equiv \vect{u}_0
-\inner{\vect{u}_0}{{\boldsymbol \nabla} S(\vect{P})}
\inner{{\boldsymbol \nabla} S(\vect{P})}{{\boldsymbol \nabla} S(\vect{P})}^{-1}
{\boldsymbol \nabla} S(\vect{P})
=(n(n+1)^{-1},-(n+1)^{-1},\ldots,-(n+1)^{-1})$ 
and 
$\vect{v}_1\equiv \vect{v}_0
-\inner{\vect{v}_0}{{\boldsymbol \nabla} S(\vect{P})}
\inner{{\boldsymbol \nabla} S(\vect{P})}{{\boldsymbol \nabla} S(\vect{P})}^{-1}
{\boldsymbol \nabla} S(\vect{P})
=(-(n+1)^{-1},\ldots,-(n+1)^{-1},n(n+1)^{-1})$. 
This leads to 
\begin{eqnarray}
\inner{\vect{u}_1}{{\boldsymbol \nabla} F}
\inner{\vect{v}_1}{{\boldsymbol \nabla} G}
=
\inner{\vect{v}_1}{{\boldsymbol \nabla} F}\inner{\vect{u}_1}{{\boldsymbol \nabla} G}\ , 
\label{u1v1}
\end{eqnarray}
from Appendix \ref{appB}. 
From Eq.~(\ref{u1v1}), we obtain 
\begin{eqnarray}
\left(
\frac{1}{P_1}-\frac{1}{P_{n+1}}\right)
\sum_{j=1}^{n+1}
\left(
\frac{1}{P_1} - \frac{1}{P_j}
\right)
\left(
\frac{1}{P_{n+1}} - \frac{1}{P_j}
\right)
=0. 
\label{u1v12}
\end{eqnarray}
Here the first factor is nonzero 
since $P_1 < P_{n+1}$ from $F(\vect{P})>0$. 
In Eq.~(\ref{u1v12}), 
since $(P_1^{-1}-P_j^{-1})\geq 0$ 
and 
$(P_{n+1}^{-1}-P_j^{-1})\leq 0$ for all $j$, 
we obtain either $P_j=P_1$ or $P_j=P_{n+1}$ for every $j$, 
which implies $P_1=\cdots =P_m < P_{m+1}=\cdots =P_{n+1}(1\leq m\leq n)$. 
We thus find that, at the local maximum, 
\begin{eqnarray}
F(\vect{P} )
&=&
P_1^{m-1}P_{n+1}^{n-m}
\left[
mP_{n+1}
\right.
\nonumber\\ 
&&
\left.
+(n+1-m)P_1-(n+1)^2P_1P_{n+1}
\right], 
\label{F}
\end{eqnarray}
and 
\begin{eqnarray}
G(\vect{P} )
=
P_1^{m}P_{n+1}^{n+1-m}\ , 
\label{G}
\end{eqnarray}
with 
\begin{eqnarray}
mP_1+(n+1-m)P_{n+1}=1\ .
\label{sumP}
\end{eqnarray}
In Appendix \ref{appD}, 
we derive 
the local maximum of $H(\vect{P} )$ 
under the constraint 
$S(\vect{P}) =1$, 
whose value is $H_m$ in Eq.~(\ref{Hm}). 
Furthermore, in Appendix \ref{appE}, we show 
the maximum of $H_m$ is given by $m=1$, 
and its value is 
\begin{eqnarray}
H_1=
\left( \frac{1}{n}\right)^n
P_1^{\U{opt}}(1-P_1^{\U{opt}})^n\ , 
\label{H1}
\end{eqnarray}
where 
\begin{eqnarray}
P_1^{\U{opt}}\equiv 
\frac{2n+3-\sqrt{4n+1}}{2(n^2+2n+2)}\ . 
\label{Popt}
\end{eqnarray}

From Eqs.~(\ref{Hlossy}) and (\ref{H1}), 
$\max_{\vect{P}\in \mathcal{R}} H(\vect{P})$ is 
equal to $\max\{ H\u{lossy}, H_1\}$. 
In Appendix \ref{appE}, 
we show $H_1 > H\u{lossy}$, 
and thus we obtain 
\begin{eqnarray}
P\u{suc}\leq 
\frac{n!(N+n)}{N}
\left( \frac{1}{n}\right)^n
P_1^{\U{opt}}(1-P_1^{\U{opt}})^n\ . 
\label{sucbound}
\end{eqnarray}

\section{Explicit construction of an optimal circuit}
\label{sec4}

\begin{figure}[t]
 \begin{center}
  \scalebox{1.0}{\includegraphics{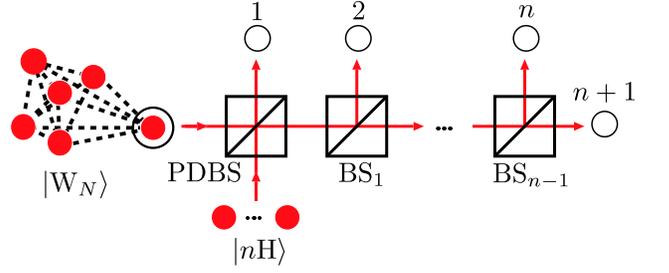}}
  \caption{(Color online) A linear optical circuit which achieves 
  the upper bound in Eq.~(\ref{sucbound}). 
  \label{fig:ExGate2}}
 \end{center}
\end{figure}
Here 
we construct 
an optical circuit which achieves 
the righthand side of Eq.~(\ref{sucbound}). 
The circuit in Fig.~\ref{fig:ExGate2} 
is composed of a PDBS and 
$n-1$ BSs, 
and this circuit has $n+1$ input and output spacial modes 
with no auxiliary spacial modes. 
We post-select the events 
where exactly one photon comes out from every output mode. 
We denote the transmittance and reflectance 
for H(V)-polarized photons of the PDBS 
by $T\u{H(V)}$ and $R\u{H(V)}$, and 
the transmittance and reflectance 
of the $k$-th BS (BS$_k$) 
by $T_{k}$ and $R_{k} (1\leq k\leq n-1)$. 
The parameters of the PDBS are set as 
$T\u{H}=R\u{V}=P_1^{\U{opt}}$, $R\u{H}=T\u{V}=1-P_1^{\U{opt}}$. 
The parameters of the BSs are set to 
output $n$ photons from $n$ BSs one by one with 
equal probability, 
that is, they are set as 
$T_{k}=(n-k)(n+1-k)^{-1}$ 
and  
$R_{k}=(n+1-k)^{-1}$ \cite{kieselDicke}. 
In this case, 
$\alpha_{1\H}=\sqrt{T\u{H}}$, 
$\alpha_{2\H}=\cdots =\alpha_{(n+1)\H}=
\sqrt{n^{-1}R\u{H}}$, 
$\beta_{j\H}=\alpha_{j\H}^{-1}-\alpha_{j\H}(n+1)$ 
and $\gamma_{j\V}=\alpha_{j\H}$ for all $j$. 
All of the other variables of 
$\alpha_{j\H(\V)}$, $\beta_{j\H(\V)}$ and $\gamma_{j\H(\V)}$ 
are equal to zero. 
We have 
$\eta_0=\cdots =\eta_{n+1}$ in this case, 
and the success probability of this circuit 
is equal to the righthand side of Eq.~(\ref{sucbound}). 
Hence we conclude that the maximum of $P\u{suc}$ is 
\begin{eqnarray}
P\u{max}\equiv 
\frac{n!(N+n)}{N}
\left( \frac{1}{n}\right)^n
P_1^{\U{opt}}(1-P_1^{\U{opt}})^n\ . 
\label{Pmax}
\end{eqnarray}
The dependence of $P\u{max}$ on $n$ and $N$ 
is shown in Fig.~\ref{fig:Pmax}. 
\begin{figure}[t]
 \begin{center}
 \scalebox{1.0}{\includegraphics{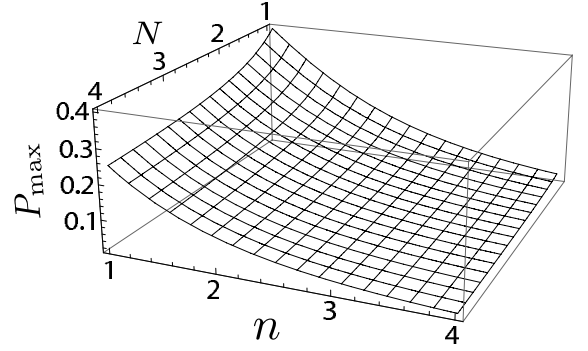}}
  \caption{The dependence of the maximum success 
  probability $P\u{max}$ in Eq.~(\ref{Pmax}) on $n$ and $N$. 
  \label{fig:Pmax}}
 \end{center}
\end{figure}

In the case of $n=1$, 
we see $P\u{max}=(N+1)/(5N)$ 
for 
$T\u{H(V)}=(5\pm\sqrt{5})/10$ and 
$T\u{V(H)}=(5\mp\sqrt{5})/10$. 
This value is the same as that in Ref.~\cite{tashima1, gong}.
In the case of $n=2$, 
we have $P\u{max}=8(N+2)/(125N)$ 
for $T\u{H}=4/5$, $T\u{V}=1/5$, $T\u{1}=1/2$ 
and $R\u{1}=1/2$. 
This value is higher than $(N+2)/(16N)$ 
given by the circuit composed of only half BSs 
in Ref.~\cite{tashima2}. 

\begin{figure}[t]
 \begin{center}
  \scalebox{1.0}{\includegraphics{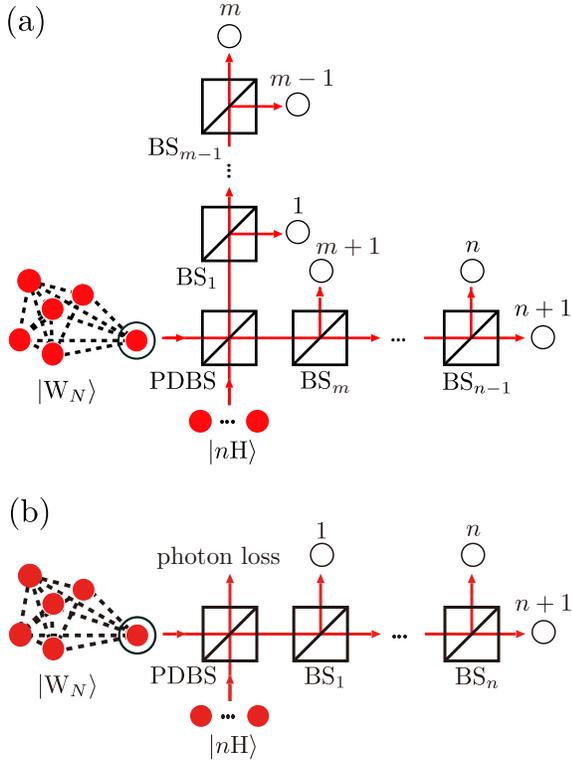}}
  \caption{(Color online) Linear optical circuits which achieve 
  (a) $P\u{suc}=n!(N+n)N^{-1}H_m$, and 
  (b) $P\u{suc}=n!(N+n)N^{-1}H\u{lossy}$. 
  }
  \label{fig:ExGate3}
 \end{center}
\end{figure}
We also construct optical circuits which achieve 
the success probability 
of the other local maximums, 
$H_m$ in Eq.~(\ref{Hm}) 
of Appendix \ref{appD} 
and 
$H\u{lossy}$ in Eq.~(\ref{Hlossy}). 
A circuit which achieves 
$P\u{suc}=n!(N+n)N^{-1}H_m$ is constructed with 
a PDBS and $n-1$ BSs. 
\{BS$_1$, $\ldots$, BS$_{m-1}$\} 
and 
\{BS$_{m}$, $\ldots$, BS$_{n+1}$\} 
are placed at each output of the PDBS in series, 
as shown in Fig.~\ref{fig:ExGate3}~(a). 
This circuit has $n+1$ input and output spacial modes 
without auxiliary spacial modes. 
Parameters of 
the 
PDBS and 
the 
BSs are set as 
$T\u{H}=R\u{V}=m\xi_m$, $R\u{H}=H\u{V}=1-m\xi_m$ 
where $\xi_m$ is in Eq.~(\ref{xim}), 
$T_{k}=(m-k)(m+1-k)^{-1}$ 
and 
$R_{k}=(m+1-k)^{-1}$ 
for $1\leq k\leq m-1$, 
and 
$T_{k}=(n-k)(n+1-k)^{-1}$ 
and 
$R_{k}=(n+1-k)^{-1}$ 
for $m\leq k\leq n-1$. 
In this case, 
$\alpha_{1\H}=\cdots =\alpha_{m\H}=\sqrt{m^{-1}T\u{H}}$, 
$\alpha_{(m+1)\H}=\cdots =\alpha_{(n+1)\H}=
\sqrt{(n+1-m)^{-1}R\u{H}}$, 
$\beta_{j\H}=\alpha_{j\H}^{-1}-\alpha_{j\H}(n+1)$ 
and $\gamma_{j\V}=\alpha_{j\H}$ for all $j$. 
All of the other variables of 
$\alpha_{j\H(\V)}$, $\beta_{j\H(\V)}$ and $\gamma_{j\H(\V)}$ 
are equal to zero. 

A circuit which achieves 
$P\u{suc}=P\u{lossy}\equiv n!(N+n)N^{-1}H\u{lossy}$ 
is constructed with 
a PDBS and $n$ BSs. 
The PDBS and BS$_1$, $\ldots$, BS$_{n}$ 
are placed in series, 
as shown in Fig.~\ref{fig:ExGate3}~(b). 
We post-select the events 
where the photons come out 
from output spacial modes of $n$ BSs, one by one. 
This circuit can be regarded as a lossless circuit with 
$n+2$ input and output modes, but if we regard it 
as a circuit with $2$ inputs and $n+1$ output, 
as in Fig.~\ref{fig:ExGate}~(a), 
it is a lossy circuit. 
Parameters of the PDBS and the BSs are set as 
$R\u{H}=1-(n+1)^{-2}$, $T\u{H}=(n+1)^{-2}$, $R\u{V}=0, T\u{V}=1$, 
$T_{k}=(n+1-k)(n+2-k)^{-1}$ 
and $R_{k}=(n+2-k)^{-1}$. 
In this case, 
$
\alpha_{1\H}=\cdots =\alpha_{(n+1)\H}=\sqrt{R\u{H}(n+1)^{-1}}$, 
$\alpha_{(n+2)\H}=\sqrt{T\u{H}}$, 
$\beta_{j\H}=\sqrt{R\u{H}}(\alpha_{j\H}^{-1}-\alpha_{j\H}(n+1))$ 
for $1\leq j \leq n+1$, 
$\beta_{(n+2)\H}=-\sqrt{R\u{H}}\alpha_{(n+2)\H}(n+1)$, 
and $\gamma_{j\V}=\sqrt{(n+1)^{-1}}$ for $1\leq j\leq n+1$. 
All of the other variables of 
$\alpha_{j\H(\V)}$, $\beta_{j\H(\V)}$ and $\gamma_{j\H(\V)}$ 
are equal to zero. 

Let us compare the success probability $P\u{lossy}$ 
for the lossy circuit with the global maximum $P\u{max}$ 
for a few examples. 
In the case of $n=1$, 
we see $P\u{lossy}=3(N+1)/(16N)$
for 
$T\u{H}=1/4$ and $T\u{1}=1/2$, 
whereas $P\u{max}=(N+1)/(5N)$. 
In the case of $n=2$, 
we have 
$P\u{lossy}=128(N+2)/(2187N)\sim 0.059(N+2)/N$ 
for $T\u{H}=1/9$, $T\u{1}=2/3$, $T\u{2}=1/2$, 
while $P\u{max}=0.064(N+2)/N$. 

\section{Conclusion}
\label{sec5}
We have derived the maximum success probability 
of the circuits composed of passive linear optics and 
an ancilla mode in an $n$-photon Fock state 
for post-selectively 
expanding an $N$-photon polarization-entangled W state 
to an $(N+n)$-photon polarization-entangled W state, 
by accessing only one photon of the initial W state. 
Whereas the symmetry in W states suggests that 
photons from beamsplitters 
should be equally distributed 
among $n+1$ output modes, 
bosonic nature of photons requires 
us to introduce either optical losses or 
to break symmetry 
in order to reduce interference effects 
between one photon from the $N$-photon W state 
and the photons from the Fock state. 
In fact, 
both cases possess local maximums 
at which the success probability does not increase 
by infinitesimal changes in variables. 
We showed that 
the overall maximum success probability is achieved 
by a PDBS and $n-1$ BSs.
In the case of $n=2$, 
the maximum success probability is higher 
than that of the expanding gate proposed in 
Ref.~\cite{tashima2}. 
 
\section*{Acknowledgements}

We thank Tsuyoshi Kitano and \c{S}ahin K. \"Ozdemir 
for helpful discussions. 
This work was supported 
by Funding Program for World-Leading Innovative R$\&$D 
on Science and Technology (FIRST), 
MEXT Grant-in-Aid for Scientific Research 
on Innovative Areas 20104003 and 21102008, 
JSPS Grant-in-Aid for Scientific Research(C) 20540389, 
and MEXT Global COE Program. 

\appendix

\section{}
\label{appA}

In the following, 
we prove that 
when $P\u{suc} > 0$ is satisfied, we have 
$\alpha_{i\V}=0$ and $\beta_{i\V}=0$ 
for $1\leq i\leq n+1$, 
and $\prod_{j=1}^{n+1}\alpha_{j\H}\neq 0$. 
For nonzero success probability $P\u{suc}>0$, 
it is necessary to have 
$\eta_0, \eta_{i} \neq 0 $. 
From Eqs.~(\ref{initial}) and (\ref{postselect}), 
the term in $\ket{\W_N}$ including $\dagg{a_{1\H}}$ 
must not be transformed 
into the terms including $\dagg{b_{i\V}}$ 
by the unitary operator $U$. 
Then we obtain 
\begin{eqnarray}
\Gamma_i &\equiv &
\bra{vac}\u{out}
\left(
b_{i\V}\prod_{j\neq i}^{n+1} b_{j\H}
\right)
U\frac{\dagg{a\u{1H}}(\dagg{a\u{2H}})^n}{n!}
\ket{vac}\u{in}\nonumber \\
&=&\beta_{i\V}\prod_{j\neq i}^{n+1}\alpha_{j\H}
+\alpha_{i\V}
\sum_{j\neq i}\beta_{j\H}
\prod_{k\neq i,j}^{n+1}\alpha_{k\H}
=0\ , 
\label{alphaH}
\end{eqnarray}
for $1\leq i\leq n+1$ and
\begin{eqnarray}
\bra{vac}\u{out}
\left(
\prod_{j=1}^{n+1}b_{j\V}
\right)
U\frac{\dagg{a\u{1H}}(\dagg{a\u{2H}})^n}{n!}
\ket{vac}\u{in}\nonumber \\
=\sum_{j=1}^{n+1}
\beta_{j\V}
\prod_{k\neq j}^{n+1}\alpha_{k\V}
=0\ . 
\label{alphaV}
\end{eqnarray}
Because 
$\sum_{i=1}^{n+1}
\Gamma_i \alpha_{i\H}\prod_{l\neq i} \alpha_{l\V}
=n\eta_0 \prod_{j=1}^{n+1}\alpha_{j\V}=0
$ is obtained from Eqs.~(\ref{alphaH}) and (\ref{alphaV}), 
we have $\prod_{j=1}^{n+1}\alpha_{j\V}=0$. 
Assuming that $\alpha_{l\V}=0$, 
we find three facts: 
(i)
$\prod_{j\neq l}^{n+1}\alpha_{j\H}\neq 0$ 
because 
$\eta_l = \gamma_{l\V}\prod_{j\neq l}^{n+1}\alpha_{j\H}\neq 0$, 
(ii)
$\alpha_{j\V}=0$ for $1\leq j\leq n+1$ 
because 
$\eta_l = \gamma_{l\V}\prod_{j\neq l}^{n+1}\alpha_{j\H}\neq 0$ 
and 
\begin{eqnarray}
\bra{vac}\u{out}
\left(
b_{l\V}b_{j\V}
\prod_{k\neq l,j}^{n+1}b_{k\H}
\right)
U\frac{\dagg{a\u{1V}}(\dagg{a\u{2H}})^n}{n!}
\ket{vac}\u{in}\nonumber\\
=\gamma_{l\V}\alpha_{j\V}\prod_{k\neq l,j}^{n+1}\alpha_{k\H}=0\ , 
\end{eqnarray}
for $1\leq j(\neq l)\leq n+1$ from Eq.~(\ref{postselect}), 
and 
(iii) $\beta_{l\V}=0$ 
from 
$\Gamma_l = \beta_{l\V}\prod_{j\neq l}^{n+1}\alpha_{j\H}=0$. 
Therefore, from 
$\prod_{j=1}^{n+1}\alpha_{j\V}=0$ and 
recursive use of (i)--(iii), 
we have $\alpha_{i\V}=0$ and $\beta_{i\V}=0$ 
for $1\leq i\leq n+1$, and 
$\prod_{j=1}^{n+1}\alpha_{j\H}\neq 0$. 

\section{}
\label{appB}

Here we prove the following statement. 
Suppose 
$H(\vect{P})\equiv \min\{ F(\vect{P}),G(\vect{P})\}$. 
Let $\epsilon > 0$ be a constant and 
$\vect{u}$ and $\vect{v}$ be arbitrary vectors. 
If 
$H(\vect{P}+\Delta \vect{P}) \leq H(\vect{P})$ 
for any 
$\Delta \vect{P}\equiv \alpha \vect{u}+\beta \vect{v}$ 
with $\alpha^2+\beta^2 < \epsilon$, 
then 
\begin{eqnarray}
\inner{\vect{u}}{{\boldsymbol \nabla} F}
\inner{\vect{v}}{{\boldsymbol \nabla} G}
=
\inner{\vect{v}}{{\boldsymbol \nabla} F}\inner{\vect{u}}{{\boldsymbol \nabla} G}\ ,
\label{uv}
\end{eqnarray}
is satisfied where 
$\inner{\vect{X}}{\vect{Y}}$ means the inner product 
between $\vect{X}$ and $\vect{Y}$. 

We define two vectors $\vect{a}$ and $\vect{b}$ as 
\begin{eqnarray}
\vect{a}\equiv 
\inner{\vect{u}}{{\boldsymbol \nabla} F}\vect{v} 
- \inner{\vect{v}}{{\boldsymbol \nabla} F}\vect{u}\ , 
\label{a}
\end{eqnarray}
and 
\begin{eqnarray}
\vect{b}\equiv 
\inner{\vect{v}}{{\boldsymbol \nabla} G}\vect{u} 
- \inner{\vect{u}}{{\boldsymbol \nabla} G}\vect{v}\ . 
\label{b}
\end{eqnarray}
Suppose 
$\inner{\vect{u}}{{\boldsymbol \nabla} F}
\inner{\vect{v}}{{\boldsymbol \nabla} G}
> 
\inner{\vect{v}}{{\boldsymbol \nabla} F}\inner{\vect{u}}{{\boldsymbol \nabla} G}$. 
From Eqs.~(\ref{a}) and (\ref{b}), we obtain 
$\inner{\vect{a}}{{\boldsymbol \nabla} F}
=\inner{\vect{b}}{{\boldsymbol \nabla} G}=0$ and 
$\inner{\vect{a}}{{\boldsymbol \nabla} G}=
\inner{\vect{b}}{{\boldsymbol \nabla} F}
> 0$. 
Hence both 
$\inner{\vect{a}+\vect{b}}{{\boldsymbol \nabla} F} > 0$ and 
$\inner{\vect{a}+\vect{b}}{{\boldsymbol \nabla} G}> 0$, 
which 
implies 
the direction of $\vect{a}+\vect{b}$ 
increases $F$ and $G$ at the same time. 
This fact contradicts with 
$H(\vect{P}+\Delta \vect{P})\leq H(\vect{P})$ for any $\Delta
\vect{P}$. 
In the case of $\inner{\vect{u}}{{\boldsymbol \nabla} F}
\inner{\vect{v}}{{\boldsymbol \nabla} G} < 
\inner{\vect{v}}{{\boldsymbol \nabla} F}\inner{\vect{u}}{{\boldsymbol \nabla} G}$, 
both $\inner{-\vect{a}-\vect{b}}{{\boldsymbol \nabla} F} > 0$ and 
$\inner{-\vect{a}-\vect{b}}{{\boldsymbol \nabla} G}> 0$ are 
satisfied, 
which also contradicts with 
$H(\vect{P}+\Delta \vect{P})\leq H(\vect{P})$ 
for any $\Delta \vect{P}$. 
Therefore we obtain Eq.~(\ref{uv}). 

\section{}
\label{appC}

We derive the maximum of 
$H(\vect{P})=\min\{ F(\vect{P}),G(\vect{P})\}$ 
in the case of 
$P_1=\cdots =P_{n+1}=S(\vect{P}) (n+1)^{-1}$, 
namely, 
\begin{eqnarray}
F(\vect{P})
=
\frac{S(\vect{P})^n}{(n+1)^{n-1}}
(1-S(\vect{P}))\ , 
\label{F2}
\end{eqnarray}
and 
\begin{eqnarray}
G(\vect{P})
=
\frac{1}{S(\vect{P})}
\left(
\frac{S(\vect{P})}{n+1}
\right)^{n+1}. 
\label{G2}
\end{eqnarray}
$F(\vect{P})$ and $G(\vect{P})$ are regarded as 
functions of single parameter $S(\vect{P})$. 
From Eq.~(\ref{F2}), 
we have 
\begin{eqnarray}
\od{F(\vect{P})}{S(\vect{P})}
=
\left(
\frac{S(\vect{P})}{n+1}
\right)^{n-1}
\left(
n-S(\vect{P})(n+1)
\right). 
\label{dF}
\end{eqnarray}
We find that 
$F(\vect{P})$ decreases monotonously 
for 
$n(n+1)^{-1} \leq S(\vect{P}) \leq 1$ 
from Eq.~(\ref{dF}), 
and 
$G(\vect{P})$ increases monotonously 
for $0<S(\vect{P})\leq 1$ from (\ref{G2}).  
From Eqs.~(\ref{F2}) and (\ref{G2}), 
the solution of $F(\vect{P})=G(\vect{P})$ is given by 
$S(\vect{P})=1-(n+1)^{-2}$. 
Since $n(n+1)^{-1} < 1-(n+1)^{-2}$ is satisfied, 
$S(\vect{P})=1-(n+1)^{-2}$ gives 
the maximum of $H(\vect{P})$, 
whose value is 
\begin{eqnarray}
H\u{lossy}\equiv \frac{n^n(n+2)^n}{(n+1)^{3n+1}}\ . 
\label{Hlossy2}
\end{eqnarray}

\section{}
\label{appD}

We derive the maximum of 
$H(\vect{P})=\min\{ F(\vect{P}),G(\vect{P})\}$ 
under Eqs.~(\ref{F}) -- (\ref{sumP}), namely, 
\begin{eqnarray}
F(\vect{P} )
=
F(\xi)
&\equiv &
\xi^{m-1}\zeta^{n-m}
\left(
m\zeta
+(n+1-m)\xi
\right.
\nonumber\\ 
&&\left.
-(n+1)^2\xi\zeta
\right), 
\label{F1}
\\
G(\vect{P} )
=
G(\xi)&\equiv &
\xi^{m}\zeta^{n+1-m}\ , 
\label{G1}
\end{eqnarray}
with 
\begin{eqnarray}
\zeta=
\frac{1-m\xi}{n+1-m}\ .
\label{zeta}
\end{eqnarray}
Because $\xi < \zeta$, 
$0 < \xi < (n+1)^{-1}$ is satisfied. 
From Eqs.~(\ref{F1}) and (\ref{G1}), 
$F(\xi)=G(\xi)$ has four roots, 
which are given by 
$\xi=0, m^{-1}$ and the two roots of 
$I(\xi)\equiv
m\zeta+(n+1-m)\xi-(n+1)^2\xi\zeta
-\xi\zeta=0$. 
Since $I(0)>0$ and $I((n+1)^{-1})<0$, 
there is only one root 
satisfying $F(\xi)=G(\xi)$ for $0<\xi<(n+1)^{-1}$, 
which is 
\begin{eqnarray}
\xi =\xi_m\equiv 
\frac{2m(n+1)+1-\sqrt{4m(n+1-m)+1}}{2m((n+1)^2+1)}.
\label{xim}
\end{eqnarray}
From Eqs.~(\ref{F1}), (\ref{G1}) and (\ref{zeta}), 
we obtain 
\begin{eqnarray}
\od{F(\xi)}{\xi}
&
=
&
m\xi^{m-2}\zeta^{n-m-1}(\zeta-\xi)f(\xi)\ ,
\label{dfdp}
\end{eqnarray}
where 
\begin{eqnarray}
\hspace{-0.4cm}
f(\xi)\equiv
\frac{m(n+1)^2}{n+1-m}\xi^2
-\frac{n+1+2nm}{n+1-m}\xi
+\frac{m-1}{n+1-m}, 
\end{eqnarray}
and 
\begin{eqnarray}
\od{G(\xi)}{\xi}
=m\xi^{m-1}\zeta^{n-m}
(\zeta-\xi)\ . 
\label{dgdp}
\end{eqnarray}
From Eq.~(\ref{dgdp}), 
we find that 
$G(\xi)$ increases monotonously for $0<\xi<(n+1)^{-1}$. 
Since 
$f(\xi)$ is a convex function, 
$f(\xi_m)=(\xi-1)(\zeta-1)-1<0$ 
and $f((n+1)^{-1})=-2(n+1)^{-1}<0$ 
assure that $F(\xi)$ decreases monotonously 
for $\xi_m \leq \xi<(n+1)^{-1}$. 
Therefore $\xi=\xi_m$ gives the maximum of $H(\vect{P})$, 
and the value is 
\begin{eqnarray}
H_m\equiv F(\xi_m)
=\xi_m^m \left(
\frac{1-m\xi_m}{n+1-m}
\right)^{n+1-m}. 
\label{Hm}
\end{eqnarray}

\section{}
\label{appE}

We show that $H_1 > H_m (2\leq m \leq n)$ 
and $H_1 > H\u{lossy}$, 
where $H_m$ is defined in Eq.~(\ref{Hm}) 
and 
$H\u{lossy}$ is in Eq.~(\ref{Hlossy}). 

Since $H_m > 0$, let us analyze the property of 
$\log H_m$ as a continuous function of $m$ 
for $1\leq m\leq n$. 
Let 
$h(m)\equiv H_m^{-1}\d{H_m}/\d{m}$ 
be its derivative. 
Because we have 
\begin{eqnarray}
\od{h(m)}{m}=\frac{(n+1)-2m}{m(n+1-m)(4m(n+1-m)+1)^{3/2}}\ , 
\end{eqnarray}
we find that 
$m=(n+1)/2$ gives the maximum of $h(m)$. 
From 
{$\d{h((n+1)/2)}/\d{n}
=2(n+1)^{-1}((n+1)^2+1)^{-3/2}> 0 $ 
for all $n$ and 
$h((n+1)/2)\rightarrow 0\ (n\rightarrow \infty)$, 
we obtain $h(m) < 0$ for all $m$.
Thus, 
because $\d{H_m}/\d{m}<0$ holds, 
$H_m$ takes its maximum for $m=1$, 
whose value is 
\begin{eqnarray}
H_1 = 
\left( \frac{1}{n}\right)^n
\xi_1(1-\xi_1)^n\ , 
\label{Hopt}
\end{eqnarray}
from Eq.~(\ref{Hm}), where $\xi_1$ is in Eq.~(\ref{xim}). 

Next we show $H_1 > H\u{lossy}$. 
Since $H_1 > H_{(n+1)/2}$ is satisfied, 
we show $H_{(n+1)/2}>H\u{lossy}$. 
From Eqs.~(\ref{Hlossy}) and (\ref{Hm}), 
\begin{eqnarray}
\frac{H_{(n+1)/2}}{H\u{lossy}}
&=&
\frac{(n+1)^4}{(n+1)^4-1}
\nonumber\\
&
&
\hspace{-1.2cm}
\times
\left(
\frac{(n+1)^4}{(n+1)^4-1}
\cdot
\frac{(n+1)^2}{(n+1)^2-1}
\right)^{\frac{n-1}{2}} 
> 1, 
\label{compH}
\end{eqnarray}
is satisfied. 
Hence 
we obtain $H_1>H\u{lossy}$.

\end{document}